\documentstyle[12pt]{article}

\def\be{\begin{equation}}
\def\en{\end{equation}}

\begin{document}
\begin{titlepage}
\baselineskip = 25pt
\begin{center}

{\Large\bf On the microwave background anisotropy produced by 
big voids in open universes}

\vspace{.5 cm}
{\bf M.J. Fullana $^{1}$, J.V. Arnau $^{2}$, \& D. S\'aez $^{1}$}\\
\small
$^{1}$ Departament d'Astronomia i Astrof\'{\i}sica. Universitat de Val\`encia.\\
46100 Burjassot (Val\`encia), Spain.\\
$^{2}$ Departament de Matem\`atica Aplicada. Universitat de Val\`encia.\\
46100 Burjassot (Val\`encia), Spain.\\
\footnotesize
e-mail: DIEGO.SAEZ@UV.ES\\

\end{center}

\vspace{2 cm}
\normalsize
\begin{abstract}

The Tolman-Bondi solution of the Einstein equations is used
in order to model the time evolution of the void observed in Bo\"otes. 
The present density contrast of the central region ($\sim -0.75$) and
its radius ($\sim 30h^{-1} \ Mpc$) are fixed, while
the density parameter of the Universe, the amplitude of the density contrast 
inside the void wall, the width of this wall and the distance from the void 
centre to the Local Group are appropriately
varied. The microwave background anisotropy produced by Bo\"otes-like
voids is estimated for a significant set of locations. All the
voids are placed far from the last scattering surface. It is shown that the 
anisotropy generated by these voids strongly depends 
on the density parameter, the wall structure and the void location.
The Doppler dipole and quadrupole are subtracted and the residual anisotropy 
is calculated. In the case of some isolated Bo\"otes-like voids 
placed at redshifts between 1 and 10 in an open universe with density 
parameter $\Omega_{0}=0.2$, the residual anisotropy
appears to be a few times $10^{-6}$ on scales of a few degrees. 
This anisotropy is about one order of magnitude greater than 
previous estimates corresponding to other cases.
The anisotropy produced by a distribution of 
voids is qualitatively studied in the light of this result.
Comparisons with previous estimates are discussed.

\end{abstract}


\end{titlepage}

\section{Introduction}

Two methods have been used in order to estimate the anisotropies
produced by the nonlinear voids of the galaxy distribution 
in {\em open universes}. 
One of these methods is based on the 
so-called {\em Swiss-Cheese} model (Rees \& Sciama 1968).
The original version of this
model applies to the case of  
overdensities surrounded by underdensities, 
but suitable modifications lead to a model for
underdensities surrounded by overdensities  
(Thompson \& Vishniac 1987). Estimates of the anisotropies
produced by Swiss-Cheese voids were obtained by Thompson
\& Vishniac (1987), and Mart\'{\i}nez-Gonz\'alez \& Sanz (1990). 
The main elements of these spherical voids are: 
an absolute vacuum in the void core, a uniform
overdense shell compensating the vacuum, 
and a general Friedmann-Robertson-Walker background outside
this shell; hence, the density profile of the
resulting structure is very particular, while the background is 
general and the compensation is exact.  The second method is
based on the Tolman-Bondi Solution (TBS) of the Einstein equations
(Tolman 1934; Bondi 1947). A general 
assymptotic Friedmann-Robertson-Walker background and
a general spherically symmetric energy density profile 
are compatible with this solution; 
such a general profile
can be used in order to model both a partial vacuum 
in the void core
and a sharped wall. 
Two different codes based on the TBS 
were built up by Panek (1992) and
Arnau et al. (1993).
Panek's code was used (Panek 1992) to 
estimate the anisotropy produced by voids with small
compensating walls,
while the code due to Arnau et al. was used (Arnau et al. 1993)
in the case of voids without walls; here, the void walls are
modeled in some detail taking into account some observational data.

In {\em flat universes}, another powerful method 
is being used in order to estimate
the anisotropy produced by nonlinear cosmological structures. 
This method does not require any symmetry, it is based on the
potential approximation developed by 
Mart\'{\i}nez-Gonz\'alez, Sanz \& Silk  (1990), which applies
beyond the linear regime. In the flat case, 
the method was applied by 
Anninos et al. (1991), Mart\'{\i}nez-Gonz\'alez,
Sanz \& Silk (1992, 1994), Tuluie \& Laguna (1995) and 
Quilis, Ib\'a\~nez \& S\'aez (1995);
these authors used various complementary techniques and conditions
(N-body simulations,
high resolution shock capturing methods and particular spectra and
statistics).

Recently, 
Arnau, Fullana \& S\'aez (1994) proved that some Great Attractor-like 
structures --evolving in open universes ($\Omega_{0}<0.4$)
and placed at redshifts between $2$ and $30$--
produce anisotropies of the order of
$10^{-5}$ on angular scales of a few degrees.
These structures had a density contrast 
of the order of 1
when they influenced the microwave photons; hence, the resulting
nonlinear gravitational anisotropies are produced  
in the mildly nonlinear regime. 
The following question arises: Are there void-like objects 
--suitable structures, locations and backgrounds-- 
producing significant anisotropies 
as in the case of the Great Attractor-like objects studied
by 
Arnau, Fullana \& S\'aez (1994)?. In the case of nonlinear structures
(density contrasts $\delta > 0.1$) evolving in open universes, 
the TBS and the Swiss-Cheese model can be used in order to answer
this question. The use of the TBS
seems to be preferable because this solution involves
appropriate density profiles. 

In order to model void-like objects, some
observational data must be taken into account.
In the eighties,
there has been a great deal of observations on the 
spatial distribution of galaxies; in these observations, 
some voids with walls have been detected (Kirshner et al. 1981,
Davis et al. 1982, Vettolani et al. 1985,
de Lapparent, Geller \& Huchra 1986, Rood 1988, 
Dey, Strauss \& Huchra 1990);
among them, the Bo\"otes Void (BV) seems to be the greatest one.
This void was first described by Kirshner et al. (1981).
From the data given by these authors and those due to 
de Lapparent, Geller \& Huchra (1986) it follows that the void in
Bo\"otes is a big quasispherical region with a defect
of galaxies; its centre is located at
$\sim$ 150 $h^{-1} \ Mpc$ from the Local Group and its 
radius is $\sim$ 30 $h^{-1} \ Mpc$. This region only contains
the $25 \%$ of the galaxies expected 
in the same volume of the background; hence,
the energy density contrast of galaxies inside the
void is $\sim -0.75$.  
Surrounding this region, there is an irregular shell having
an excess of galaxies, 
in other words, there is an inhomogeneous
sharped wall. 
Although the observations are not very accurate 
(Kirshner et al. 1981,
de Lapparent, Geller \& Huchra 1986), current data 
suggest that the amplitude of the density contrast of
galaxies inside the wall is $\sim 4$ and the mean width
of this wall is $\sim 5h^{-1} \ Mpc$. 

The above observational data must be complemented with 
appropriate assumptions about the dark matter distribution.
Since the nonlinear anisotropy produced by voids located far from the 
last scattering surface is a gravitational effect, this anisotropy 
is produced by the total energy density contrast
$\frac {\Delta \rho}{\rho}$. This contrast can be obtained from 
the observational value  of the density contrast produced by galaxies 
$(\frac {\Delta \rho}{\rho})_{gal}$ and 
the value of the the so-called
{\em linear bias parameter b}. 
This parameter is defined by the relation
$(\frac {\Delta \rho}{\rho})_{gal} = b (\frac {\Delta \rho}{\rho})$.
In this paper, it is assumed that luminous
galaxies trace the mass distribution. This means that the parameter $b$
is assumed to be unity and, consequently,  
the relative defects (excesses) of galaxies 
and dark matter are identical inside the 
void (wall). The case $b < 1$ has not either theoretical or observational
support, it corresponds to a void (wall)  with an amount of dark matter
smaller (greater) than that of the case $b = 1$; by this reason, 
the anisotropy of the case $b < 1$ is expected to be greater than
that estimated in the case $b = 1$. Similar arguments lead to the
conclusion that, in the case $b > 1$, the anisotropies are smaller
than those of the case $b=1$. Since the condition $b = 1$ is used along 
the paper, we can state that our computations give upper limits to
the anisotropy produced by Bo\"otes-like objects, except in the
unlikely case $b < 1$.

In this paper, the wall is described by two parameters, the
amplitude, $(\frac {\Delta \rho}{\rho})_{max}$, 
of the total density contrast inside the wall 
and the distance, $d_{w}$, between the two points of
the wall in which $\frac {\Delta \rho}{\rho}$ is the $20 \%$ 
of $(\frac {\Delta \rho}{\rho})_{max}$; the distance $d_{w}$
is called the wall width. 
In the case $b=1$,
current data suggest that
the value of $(\frac {\Delta \rho}{\rho})_{max}$ 
is  $\sim 4$ and
the wall width $d_{w}$ is $\sim 5h^{-1} \ Mpc$.
It can be easily verified (see Section 2) that
a wall having these features overcompensates the
central underdensity described above. The mass excess of this wall 
is about two times the mass defect of the underdense 
region. This is not a model dependent conclusion, but 
a direct consequence of
the observations. 
In this task, the anisotropy produced by an isolated structure
is estimated. The chosen overcompensated structure is formed by a void
and all the matter surrounding it.
According to 
the above observational evidences, overcompensated 
structures of this kind are
present in the universe. As required by
the cosmological principle, each of these structures should be
compensated 
by other structures 
in large volumes containing various voids. In order to imagine
this compensation, it is useful to 
take into account that
the energy excess 
surrounding a certain underdensity 
is shared by the neighboring ones and, consequently, this excess
also contributes to the 
compensation of other neighboring underdensities; in other words,
only a part of this excess must compensate the central underdensity.

The anisotropy produced by a realistic distribution of 
irregular voids and walls cannot be calculated from the 
anisotropy produced by an isolated void with walls; nevertheless,
if the chosen isolated structure produces 
large enough anisotropies and its spatial distribution is
appropriate,   
the true distribution of voids could 
produce relevant effects; in this case, 
the information obtained from the study of isolated 
structures strongly motivates further researches based on
suitable approaches.
Since the effects of isolated voids with walls 
placed far from the last scattering surface are expected to be
small, 
the greatest observed void, namely, the BV has been selected. 
Several
BV realizations evolving in various backgrounds
have been considered in order to
do an exhaustive study of the anisotropies
produced by isolated voids. Observational evidences are 
taken into account in order to select these realizations and 
backgrounds. 

Henceforth, $a$ is the scale factor, $t$ is the cosmological time,
an overdot stands for a derivative with respect to $t$, $H$ is the ratio
$\dot{a}/a$, and $\Omega_{0}$ is the density parameter. The subscripts $D$,
$0$ and $B$ indicate that a quantity has been valued 
at decoupling, at present time and in the background, respectively;
for instance, $H_{0}$ is the Hubble constant.
If $H_{0}$ is given 
in units of $km \ s^{-1} \ Mpc^{-1}$, the parameter $h = H_{0}/100$
is the reduced dimensionless Hubble constant. 

The plan of this paper is as follows: 
Several BV models compatible with
current observations are defined in Section 2. 
The initial conditions --at decoupling-- leading to
these models are derived in Section 3.  
The anisotropies produced by the
BV models of Section 2 are presented in Section 4; 
appropriate comparisons with previous computations are also given in 
this section.
Finally, the main conclusions are summarized
and discussed in Section 5.

\section{BV models}

A BV realization is defined by the present 
density contrast inside the void 
$(\frac {\Delta \rho}{\rho})_{v}$, the present radius of the underdense
region $R_{v}$, the present amplitude of the density contrast inside
the wall
$(\frac {\Delta \rho}{\rho})_{max}$ and the present wall width
$d_{w}$. Any present configuration of the void in
Bo\"otes is a
BV realization.  
Each BV realization is the final state of an evolutionary 
process, which takes place in a certain 
Friedmann-Robertson-Walker Universe.
Hereafter, a BV realization and a background 
define a BV model.
Any model describes the time evolution of the void in a certain 
background. This evolution leads to a present state (a realization).
If the cosmological constant vanishes, 
$h$ and $\Omega_{0}$ are the free parameters of the background.
The numerical codes need a fixed value of $h$; nevertheless,
if the distances are given in units of $h^{-1} \ Mpc$, 
the final results do not depend on
the chosen value of $h$; so, only the background parameter
$\Omega_{0}$ is a physically significant free 
parameter to be varied.

Six BV realizations have been selected to be 
considered in the next sections.

One of the chosen BV realizations corresponds to  
$(\frac {\Delta \rho}{\rho})_{v} \sim -1$,
$R_{v} \sim  30h^{-1} \ Mpc$, 
$(\frac {\Delta \rho}{\rho})_{max} \sim 2.5$ and 
$d_{w} \sim 2h^{-1} \ Mpc$. This realization is only 
studied in the case $\Omega_{0}=1$. The
resulting model is considered with the essential aim of testing our
codes and comparing our results with previous ones 
(Panek 1992)

The remaining five BV realizations are obtained as follows: 

The quantities  
$(\frac {\Delta \rho}{\rho})_{v}$ and $R_{v}$ 
are fixed; their values are assumed to be $\sim -0.75$ and
$30h^{-1} \ Mpc$, respectively.
The amplitude $(\frac {\Delta \rho}{\rho})_{max}$ is varied
from 2 to 6, and the wall width $d_{w}$ is varied from 
$3h^{-1} \ Mpc$ to $7h^{-1} \ Mpc$.   
Taking into account that the
values of these parameters suggested by
the observations (de Lapparent, Geller \& Huchra 1986) are 
$(\frac {\Delta \rho}{\rho})_{max} \sim 4$ 
and $d_{w} \sim 5h^{-1} \ Mpc$, we proceed as follows: 
in a first step, the wall width $d_{w}=5h^{-1} \ Mpc$
is fixed and the amplitudes
$2$, $4$ and $6$ are considered and, in a second step, the
amplitude $(\frac {\Delta \rho}{\rho})_{max}=4$
is fixed and 
the wall widths are assumed to be $3h^{-1} \ Mpc$, $5h^{-1} \ Mpc$
and $7h^{-1} \ Mpc$. Note that the realization 
$(\frac {\Delta \rho}{\rho})_{max} \sim 4$ 
and $d_{w} \sim 5h^{-1} \ Mpc$ is considered in each of the above steps. 
Hereafter, any of these five realizations is identified
by the values of the quantities 
$(\frac {\Delta \rho}{\rho})_{max}$ and $d_{w}$. 
Each of these  realizations
has been studied in several cases corresponding to
$\Omega_{0}$ values ranging from $0.2$ to $1$; 
nevertheless, for
the sake of briefness, only some appropriate models corresponding
to $\Omega_{0}=0.2$ and $\Omega_{0}=1$ are presented.

For each BV realization,
the wall compensates the central underdensity at a certain 
distance from the
void centre. This distance is called the compensation radius, $R_{c}$.
Tables 1 and 2 (seventh column) show these radius 
for all the models considered in this
paper. In the cases of the first and fourth rows of Table 1 and 
the second row of Table 2, the compensation occurs outside the wall,
while in the remaining cases, it occurs
inside the wall; for example, in the case
$(\frac {\Delta \rho}{\rho})_{max} \sim 4$ 
and $d_{w} \sim 5h^{-1} \ Mpc$, the compensation radius is
$R_{c}=32.7h^{-1} \ Mpc$ and, consequently, 
the compensation of the central underdensity takes place 
near the wall centre; this means that about one-half of the
wall compensates the central void, while the remaining of the wall
must compensate other underdensities. 

For a given $\Omega_{0}$ value, the quantities defining 
any BV realization (present contrasts and distances) 
must be numerically obtained 
--after evolution-- from appropriated initial conditions;
the choice of the initial conditions corresponding to a given model
(a BV realization evolving in a defined background) 
is now discussed. 

\section{Initial conditions}

The main goal of this task is the estimation of the secondary
gravitational anisotropies produced by big voids
located far from the last scattering surface. Since the evolution
of the microwave photons must be studied 
from the last scattering surface to
our position in the Universe, the initial conditions 
for the void evolution 
are set at decoupling time (redshift $z_{dec} = 1000$). 

The initial profiles of the total energy density and the
peculiar velocity field  fix the two arbitrary functions
involved in the TBS (see Arnau et al. 1993);
hence, these initial profiles fix 
the time evolution of the resulting void from 
decoupling to present time.  

In this paper, the initial density profile is assumed to be  

\begin{equation}
\rho = \rho_{_{BD}} \left[ 1 + \frac{\varepsilon_{1}}{1 +
\left( R / R_{x1} \right)^{6}} + \frac{\varepsilon_{2}}{1 +
\left( R / R_{x2} \right)^{6}} \right] \ ,
\end{equation}
\\
where $R$ is a radial coordinate and the
conditions $\varepsilon_{1} > 0, \varepsilon_{2} < 0,
| \varepsilon_{1} | < | \varepsilon_{2} |$ and $R_{x1} > R_{2x}$
are satisfied.

The initial peculiar velocity is 
\begin{equation}
V_{_{D}} = - \frac{1}{3} H_{_{D}} 
R \left\langle \frac{\rho - \rho_{_{BD}}}{\rho_{_{BD}}}
\right\rangle \Omega_{_{D}}^{0.6}
\end{equation}
\\
where the angular brackets denote a mean value from $R = 0$ to $R$.

Since the cosmological constant vanishes,
the background parameters  
involved in Eqs. (1) and (2) can be
written in terms of $\Omega_{0}$ and $h$. 

Only the choice of the profile (1) is arbitrary. The profile 
(2) is obtained from Eq. (1). It corresponds to vanishing
nongrowing modes (Peebles 1980).
The form of the initial density profile (1) has not any theoretical 
justification. This form only gives a certain parametrization of 
the initial conditions, this parametrization  
is expected to be suitable in order to
describe voids with walls as a result of two facts: (1) 
for small values of $R/R_{x2}$, 
the quantities $(R/R_{x1})^{6}$ and $(R/R_{x2})^{6}$
become very small and $\rho$ becomes quasiconstant; this means that 
the profile (1) describes a central underdensity with
a density contrast $\sim \varepsilon_{1} + \varepsilon_{2}<0$ and, 
(2) the central underdensity is initially surrounded by
an overdensity, which is the origin of the present void wall.
It has been verified that the exponent $6$ is suitable in order to
get the required amplitudes and wall widths at present time,
but other exponents could be also tested.

A realization is defined by the quantities 
$(\frac {\Delta \rho}{\rho})_{v}$, $R_{v}$,
$(\frac {\Delta \rho}{\rho})_{max}$ and 
$d_{w}$. 
The question is: 
which are the values of the parameters
$\varepsilon_{1}$, $\varepsilon_{2}$, $R_{x1}$ and $R_{x2}$
leading to a given realization in a fixed
background?. 

For each $\Omega_{0}$ value,
a numerical code based on the TBS plus Eqs. (1)
and (2) calculates the quantities 
$(\frac {\Delta \rho}{\rho})_{v}$, $R_{v}$,
$(\frac {\Delta \rho}{\rho})_{max}$ and 
$d_{w}$ from initial values of
$\varepsilon_{1}$, $\varepsilon_{2}$, $R_{x1}$ and $R_{x2}$;
this means that the quantities defining a BV realization
are not initial conditions for the numerical code, but quantities 
derived from it.   

Given a BV model, the corresponding initial conditions
are obtained as follows: 
arbitrary values of
$\varepsilon_{1}$, $\varepsilon_{2}$, $R_{x1}$ and $R_{x2}$
are assumed and the resulting values of 
$(\frac {\Delta \rho}{\rho})_{v}$, $R_{v}$,
$(\frac {\Delta \rho}{\rho})_{max}$ and
$d_{w}$ are compared with those of the chosen model; 
if they are different, the parameters
$\varepsilon_{1}$, $\varepsilon_{2}$, $R_{x1}$ and $R_{x2}$
are varied and the results are compared again.
These calculations and comparisons are carried out 
by a numerical code based on the   
"gradient method" ( as in S\'aez, Arnau \& Fullana 1993).  
This code modifies the
initial values of the parameters
$\varepsilon_{1}$, $\varepsilon_{2}$, $R_{x1}$ and $R_{x2}$ in
such a way that the new parameters lead to
a BV model better than the previous one. 
This code
repeats the modification of the parameters 
until the resulting BV model is similar enough to
the required one. This process requires nonlinear 
techniques because the final BV model is a nonlinear
one (see Fig. 1); in other words, nonlinear methods are necessary
as a result of our BV normalization, which is based on present nonlinear
observational data.

Table 1 gives the initial values of the parameters 
$\varepsilon_{1}$, $\varepsilon_{2}$, $R_{x1}$ and $R_{x2}$ 
for each of the 
$\Omega_{0} = 0.2$ models studied in this paper. 
Table 2 gives the same information for the
$\Omega_{0} = 1$ models.  
The present $\frac {\Delta \rho}{\rho}$
profiles corresponding to  
our BV models are shown in Fig. 1. For a given realization,
the present energy density
profiles of the 
models $\Omega_{0}=0.2$ and $\Omega_{0}=1$ are indistinguishable because
the same values of 
$(\frac {\Delta \rho}{\rho})_{v}$, $R_{v}$,
$(\frac {\Delta \rho}{\rho})_{max}$ and
$d_{w}$ have been chosen in both cases (see Tables 1 and 2).

It is well known that the TBS only
applies before shell crossing. 
Hellaby and Lake (1985) gave the necessary and
sufficient conditions for 
the presence of shell crossing; these conditions are
satisfied in the cases studied in this paper; hence,
the shell crossing is unavoidable. The time at which
this phenomenon takes place is not given by the 
Hellaby and Lake conditions; this time must be determined in each
particular case.
In the cases studied in this paper,
it has been verified that the shell crossing does not
take place before present time; hence, the TBS
can be used in our computations.

\section{Anisotropy}

The initial conditions discussed in Sect. 3
define the space-time structure and, consequently,
these conditions fix the differential equations 
of the photon trajectories. 
These differential equations must be integrated 
in order to estimate the anisotropy produced by a 
BV model;  
this integration is carried out by
using the code due to Arnau et al. (1993) and
S\'aez, Arnau \& Fullana (1993) plus the initial profiles (1) and (2).

The centre of the BV open models ($\Omega_{0}=0.2$)
has been located at a significant set of
distances from the observer, which correspond to redshifts
between 0.052 ($\sim 150 h^{-1} \ Mpc$) and 100 ($8360h^{-1} \ Mpc$); 
nevertheless, only the results corresponding to
a few appropriate distances are displayed in the Figures.
The centre of the BV flat model corresponding to
the first row of Table 2 has been placed at 
the same redshifts, while the model used for comparisons with
previous computations (second row of Table 2) has been only located at 
$100 h^{-1} \ Mpc$ in order to facilitate these
comparisons; hence, all the selected structures
are located far from the last scattering surface and, 
consequently, they produce negligible
temperature fluctuations and Doppler shifts on this 
surface; in which,
the temperature is assumed to be constant.

Our code (Arnau et al. 1993, S\'aez, Arnau \& Fullana 1993)
numerically
computes the temperature T of the microwave background 
as a function of the
observation angle $\psi$; this is the angle formed by the 
line of sight and the line
joining the observer and the inhomogeneity centre.
The function $T(\psi)$ is then used to calculate the mean temperature
$<T> \ = \ (1/2) \ \int^{\pi}_{0} T(\psi) \ sin\psi \ d\psi$
and the total temperature
contrast $\delta_{T}(\psi) \ = \ [T(\psi) \ - \ <T>] \ / \ <T>$.
In the expansion of $\delta_{T}$ in spherical harmonics,
$\delta_{T}(\psi) \ = \ D \ cos \psi \ + \ Q \ (3 cos^{2} \ \psi \ - \ 1) \
+ \ higher \ order \ multipoles$, $D$ and $Q$ are the total
dipole and quadrupole, respectively. The dipole $D$ is assumed to be
a Doppler effect appearing as a result of 
the present peculiar velocity of the observer produced by the 
Bo\"otes-like object; in
other words, any gravitational contribution to the dipole is neglected; thus
the relativistic Doppler quadrupole is $D^{2}/3$. The total Doppler effect
(dipole and quadrupole) produced by the peculiar 
motion of the observer is subtracted
from $\delta_{T}(\psi)$ to obtain the residual anisotropy
$\delta_{R}(\psi) \ = \ \delta_{T}(\psi) \ - \ D \ cos \psi \ -
\ D^{2}/3 \ (3 cos^{2} \ \psi \ - \ 1)$; therefore, on account
of the large distance separating the chosen voids from the
last scattering surface,
this anisotropy is a pure gravitational effect. The rigorous
computation of
this effect requires nonlinear techniques when the 
amplitude of the density contrast reaches values greater than
$\sim 0.1$ in
some region of the structure.

The residual anisotropies produced by the BV models
of Sect. 2 --for appropriate locations-- are now presented and discussed. 

\subsection{Open Universe, $\Omega_{0} = 0.2$}

The upper panels of Fig. 2 show the residual anisotropy 
produced by three BV models. The background is open
($\Omega_{0}=0.2$), the 
wall width is $d_{w} = 5h^{-1} \ Mpc$ in all the cases, and 
the values of the amplitude $(\frac {\Delta \rho}{\rho})_{max}$
are 2, 4 and 6. The present energy density contrasts of
these models are presented in the upper panel of Fig. 1.
The initial values of the free parameters are 
given in Table 1.
In the upper left (right) panel of Fig. 2, the void is centred 
at $z=0.0052$ ($z=2.42$). The first of these redshifts 
corresponds to the 
location of the true BV and the second one to the position
leading to the maximum anisotropy (see below).
As it is observed in the plot, the
greater $(\frac {\Delta \rho}{\rho})_{max}$, 
the greater the amplitude of $\delta_{_{R}}$. In the upper left
(right) panel, the
maximum amplitude of the residual anisotropy is
$\delta_{_{R}} \sim 8 \times 10^{-7}$
($\delta_{_{R}} \sim 4.5 \times 10^{-6}$); this value is obtained in the
case $(\frac {\Delta \rho}{\rho})_{max}= 6$.
For the model $d_{w} = 5h^{-1} \ Mpc$, $(\frac {\Delta \rho}{\rho})_{max}=4$, 
the residual anisotropies are 
$\delta_{_{R}} \sim 4 \times 10^{-7}$ (left) and
$\delta_{_{R}} \sim 2.3 \times 10^{-6}$ (right).

The bottom panels of Fig. 2 also display the residual anisotropy
produced by three BV models. The background is the same
as in the upper panels, but the realizations are different.
The wall widths,
are $3h^{-1} \ Mpc$, $5h^{-1} \ Mpc$ and $7h^{-1} \ Mpc$ and 
the amplitude is $(\frac {\Delta \rho}{\rho})_{max}=4$ in
all the cases. The present energy density contrasts 
of these models are given in 
the intermediate panel of Fig. 1 and the initial conditions
can be found in Table 1.
The redshifts of the void centres are the same as in the
top panels: $z=0.052$ (left) and $z=2.42$ (right).
In the bottom left (right) panel, the  maximum
amplitude of the residual anisotropy is 
$\delta_{_{R}} \sim  6.8 \times 10^{-7}$
($\delta_{_{R}} \sim  3.8 \times 10^{-6}$). These values correspond to
the maximum wall width $d_{w}=7h^{-1} \ Mpc$. 
The greater $d_{w}$, the greater the amplitude of 
$\delta_{_{R}}$.

All the cases of the left panels of Fig. 2 correspond to the same 
backgrounds and locations and 
the central underdense regions have the same structure; hence,
the differences between the anisotropies of
two of these cases are due to the wall. 
The same can be stated for the
cases of the right panels.
In the cases corresponding to the continuous and dashed lines, 
the anisotropy produced by
the wall dominates 
the total effect. 

The upper panel of Fig. 3 shows the residual anisotropy
produced by the model  $\Omega_{0}=0.2$,
$(\frac {\Delta \rho}{\rho})_{max}=4$, and
$d_{w}=5h^{-1} \ Mpc$.
Each curve corresponds to a location of
the void. The redshift defining this location is given
inside the panel. This model is also considered in Fig. 2.
As it is shown in this panel, the amplitude of the residual anisotropy
is an increasing function of the redshift $z$ --defining the 
location of the void-- from $z=0.052$ to $z \simeq 2.42$, while it
becomes a decreasing function for $z>2.42$. The
maximum amplitude is $2.36 \times 10^{-6}$, it is 
is found at $z \simeq 2.42$. As it is 
pointed out below, this behavior is not observed in the
case of flat models. 

Fig. 4 shows the density profiles of the three realizations described 
in the top panel of Fig. 1, for $z=2.42$ and $\Omega_{0}=0.2$. 
As it can be seen in Fig. 4,
these structures were evolving
in the mildly nonlinear regime when they produced the maximum
anisotropy. The amplitude of the density contrast inside 
the underdensity is $-0.59$ in all the cases. Inside the wall,
this amplitude takes on the values $0.48$, $0.69$ and $0.82$
for the amplitudes $2$, $4$ and $6$, respectively; hence, 
the standard Eulerian linear approach does not apply. 

\subsection{Flat Universe, $\Omega_{0} = 1$}

For $\Omega_{0}=1$, the residual anisotropy corresponding to the
realization $(\frac {\Delta \rho}{\rho})_{max}=4$,
$d_{w}=5h^{-1} \ Mpc$ is shown in the intermediate panel of Fig. 3.
The void centre is located at the redshifts displayed 
inside the panel. 
The initial conditions are given in Table 2.
For a flat background, the amplitude of the 
anisotropy produced by the chosen BV realization
(the same as in the top panel of Fig. 3) is
$\sim -3.2 \times 10^{-7}$ in the case $z=0.052$.
The modulus of this amplitude decreases as $z$ increases from
$z=0.052$ to $z \simeq 2.42$ and it is an slowly increasing function of 
z for $z>2.42$;
hence, at redshifts
between  $1$ and $10$, the value of this modulus is much	
smaller than the amplitude of the residual anisotropy corresponding
to the open case (top panel of Fig. 3).
At $z=2.42$, the ratio between this amplitude and the
mentioned modulus is $\sim 8$; therefore, we can state that,
at low redshifts, the anisotropies corresponding to the open
case are much greater than those of the flat case.

In the intermediate panel of Fig. 3, it can be seen that
the wall produces a local effect.
When the structure is located at $z=0.052$,
this effect appears between $\psi=8^{\circ}$ and
$\psi=13^{\circ}$. As the redshift increases, the
effect appears at smaller angles. In any case, the angular position 
of the feature coincides with that of the sharped wall.
The photons coming along these directions cross a great part of the
wall. The accuracy of our codes allows us to obtain these small features.

The
sign of $\delta_{_{R}}(\psi=0)$ is positive for
$\Omega_{0}=0.2$ and negative for $\Omega_{0}=1$
(see the upper and intermediate panels of Fig. 3). In the absence of walls
(Arnau et al. 1993) as well as in the case of small compensating walls
(Panek 1992), the sign of $\delta_{_{R}}( \psi = 0 )$ is negative for 
both $\Omega_{0}$ values; therefore the sign change only appears
in the case of overcompensated voids.

\subsection{Comparisons with previous calculations}

Panek (1992) studied four void realizations evolving in a
flat background. The model (c) of Panek's
paper is a BV model having the following
features: $\Omega_{0}=1$,
$(\frac {\Delta \rho}{\rho})_{v} \sim -1$, $R_{v} \sim 30h^{-1} \ Mpc$,
$(\frac {\Delta \rho}{\rho})_{max} \sim 2.5$ and
$d_{w} \sim 2h^{-1} \ Mpc$.  
Our code --based on the gradient
method-- has been used in order to find the initial conditions
corresponding to this model. The values of 
$\varepsilon_{1}$, $\varepsilon_{2}$, $R_{x1}$ and $R_{x2}$
are given in Table 2 (second row). 
The present density contrast of
this model is displayed in the bottom panel of
Fig. 1.
It has been verified that the wall compensates the
central underdensity at $R_{c}=51.6h^{-1} \ Mpc$.
As in Panek's paper, the void centre is placed at
$100h^{-1} \ Mpc$ from the observer in order to compute anisotropies.
The residual anisotropy produced
by this model is plotted in the bottom panel of Fig. 3.
This panel and the bottom panel of Fig. 1  
are to be compared with
Figs. 6 and 1 of Panek's paper, respectively.
These panels have a special format --different from that of
the remaining ones-- in order to facilitate comparisons with
Panek's Figures (Panek 1992).
These comparisons clearly
show that our codes --based on the TBS
and Eq. (1) and (2)-- have let to a BV model
very similar to that of Panek (1992); accordingly, 
the residual anisotropy 
appears to be very similar  to that  predicted by this author in
his case (c). These results simultaneously test our codes and
those used by Panek (1992).

Thompson \& Vishniac (1987) and Mart\'{\i}nez-Gonzalez \& Sanz
(1990) predicted BV anisotropies  of a few times $\sim 10^{-7}$ 
for Bo\"otes-like objects located at $z \sim 0.05$ in
any admissible background and similar anisotropies for
other redshifts in a flat universe. These authors used
the Swiss-Cheese model. 
In the case of small compensating walls (bottom panel of
Fig. 3) and in the absence
of walls (Arnau et al. 1993),
our results essentially
agree with these previous estimates; 
however, under the following assumptions: (1) overcompensating
walls with the features suggested by the observational data, (2)
an open universe with $\Omega_{0}=0.2$, and (3) low redshifts 
ranging in the interval (1,10), previous predictions are 
magnified by a factor $\sim 10$. 

\section{Conclusions and discussion}

Equation (1) defines a good parametrization of the initial
density profiles in the case of 
voids with sharped walls. Our code based on the gradient
method --plus Eqs (1) and (2)-- gives the initial 
conditions leading to any BV model. 

As a result of the fact that the void walls have been 
modeled taking into account the
observational evidences, the compensation of the central underdensity
takes place at scales
larger than that of a single void. Various voids contribute
to this compensation.

The anisotropy produced by a Bo\"otes-like void strongly depends on
the wall structure, the density parameter and the location
of the symmetry centre. According to previous estimates, 
which are confirmed in this paper, the anisotropy produced
by compensated voids evolving in a flat universe has an
amplitude of a few times $10^{-7}$; however, for $\Omega_{0}=0.2$
and locations between $z=1$ and $z=10$, the overcompensated voids
suggested by the observations produce anisotropies of a few
times $10^{-6}$ on scales of a few degrees. A question is relevant:
What is the effect produced by a distribution of voids in an 
open universe?.
In the flat case $\Omega_{0}=1$, 
the anisotropies corresponding to the same range of redshifts
are much smaller. 
The value of the density parameter is of crucial importance.
In the case of overcompensating walls,
the sign of the effect towards the central region of the 
void is positive (negative) for open (flat) universes.
If this effect is detected in future in the case of a single
observable structure, results could be used in order
to constraint the density parameter; nevertheless, it should be 
pointed out that such a detection is not easy, in particular,
in the flat case, where the resulting anisotropy is very small.

For the above interval of redshifts $(1, 10)$ and $\Omega_{0}=0.2$,   
the present distances from the 
void centre to the observer range from $\sim 2000h^{-1} \ Mpc$
to $\sim 5900h^{-1} \ Mpc$. There are many
voids located between these distances; nevertheless, only some
rare voids would be Bo\"otes-like voids (or greater) 
producing anisotropies of a few times $10^{-6}$. Given two observation
angles $\psi_{1}$ and $\psi_{2}$, the number of big voids $n_{1}$ and
$n_{2}$ crossed by the photons traveling along the chosen
directions can be different. For $\mid n_{1}-n_{2} \mid > 1$,
the relative temperature difference
corresponding to $\psi_{1}$ and $\psi_{2}$ would be near $10^{-5}$.
This possibility cannot be rejected {\em a priori}. It must
be either rejected or accepted
after quantitative calculations. The feasibility of the condition
$\mid n_{1}-n_{2} \mid > 1$ depends on the abundance of big 
voids.
Since the anisotropy --on scales of a
few degrees-- observed 
in experiments as COBE (Smoot et al. 1992) and Tenerife (Watson
et al 1992) is near $10^{-5}$, the contribution of 
big voids at low redshifts could be important. 
In the flat case, this contribution is expected to be too small.
In any case, it would appear 
superimposed to the primary anisotropy produced near
the last scattering surface in the linear regime.  

Similar results were obtained in the case of
Great Attractor-like objects 
(Arnau, Fullana \& S\'aez 1994). For $\Omega_{0}<0.4$ and 
$2<Z<30$, these structures
produce anisotropies of the order of $10^{-5}$ on scales of a few
degrees. In both cases, either the universe is
open enough or the anisotropy is negligible. 
Results about Great Attractor-like object enhance the interest of
the above question, which should be rewritten as follows:
What is the anisotropy produced by a distribution of voids and
great overdensities in an open universe?.

Although a model of overcompensated isolated voids 
based on the TBS is currently  
competitive, it has some 
important limitations related to the spherical symmetry. Even if
the central underdense region is quasispherical,
the true wall is not regular and the
motion of the matter contained in this part of the structure  
is not strictly radial.
There are clusters and structures in the walls, which
produce local peculiar motions tangent to the wall. These
motions would also produce anisotropy 
(Tuluie \& Laguna 1995). 
The anisotropy produced by the substructures of the
walls are expected to be important on angular scales
smaller than a few degrees and, consequently, the
estimates of this paper should be admissible.

\vspace{0.5 cm}
{\bf ACKNOWLEDGMENTS}

This work has been supported by the project GV-2207/94.
The numerical computations have been carried out in 
the Computational Center of the
" Universitat de Val\`encia ". One of us, M.J.F., gives thanks to the
"Conselleria de Cultura, Educaci\'o i Ci\`encia de la Generalitat
Valenciana" for a fellowship.

\vspace{0.5 cm}
{\Large\bf References}
\\
Anninos, P., Matzner, R.A., Tuluie, R.
\& Centrella, J. 1991, ApJ, 382, 71\\ 
Arnau, J.V., Fullana, M.J., Monreal, L. \& S\'aez, D. 1993, ApJ,
402, 359\\
Arnau, J.V., Fullana, M.J. \& S\'aez, D. 1994, MNRAS, 268, L17\\
Bondi, H. 1947, MNRAS, 107, 410\\
Bouchet, F.R., Juszkiewicz, R., Colombi, S. \&  Pellat, R., 1992,
394, L5\\
Davis, M., Huchra, J., Latham, D.W. \& Tonry, J., 1982, ApJ,
253, 423\\
de Lapparent, V., Geller, M.J. \& Huchra, J.P. 1986, ApJ, 302, L1\\
Dey, A., Strauss, M.A., \& Huchra, J., 1990, ApJ, 99, 463\\
Hellaby, C. \& Lake, K. 1985, ApJ, 290, 381\\
Kirshner, R.P., Oemler, A.Jr., Schechter, P.L. \& Shectman, S.A. 1981,
ApJ, 248, L57\\
Mart\'{\i}nez-Gonz\'alez, E., \& Sanz, J.L. 1990, MNRAS, 247, 473\\
Mart\'{\i}nez-Gonz\'alez, E., Sanz, J.L., \& Silk, J.,
1990, ApJ, 355, L5\\
Mart\'{\i}nez-Gonz\'alez, E., Sanz, J.L., \& Silk, J.,
1992, Phys. rev., 46D, 4193\\
Mart\'{\i}nez-Gonz\'alez, E., Sanz, J.L., \& Silk, J.,
1994, ApJ, 436, 1\\
Panek, M. 1992, ApJ, 388, 225\\ 
Peebles, P.J.E., 1980, The Large Scale Structure of the Universe,
Princeton Univ. Press, Princeton, NJ\\
Quilis, V., Ib\'a\~nez, J.M., \& S\'aez, D., 1995, MNRAS, in press\\
Rees, M., \& Sciama, D.W. 1968, Nature, 217, 511\\
Rood, H.J., 1988, Ann. Rev. Astron. Astrophys., 26, 245\\
S\'aez, D., Arnau, J.V. \& Fullana, M.J. 1993, MNRAS, 263, 681\\
Smoot, G.F. et al., 1992, ApJ, 396, L1\\
Thompson, K.L. \& Vishniac, E.T. 1987, ApJ, 313, 517\\
Tolman, R.C. 1934, Proc. Natl. Acad. Sci., 20, 169\\
Tuluie, R., \& Laguna, P., 1995, ApJ, 445, L73\\
Vettolani, G., de Souza, R.E., Marano, B., \& Chincarini, G., 
1985, Astron. Astrophys., 144, 506\\
Watson, R.A., Gutierrez de la Cruz, R.D., Davies, R.D., Lasenby,
A.N., Rebolo, R., Beckman J.E., \& Hancock S., 1992, Nat, 357, 660\\

\begin{center}
{\bf Figure Captions}
\end{center}

\noindent
{\bf Fig. 1.} Present density contrast $\Delta\rho/\rho(t_{0})$
as a function of the present radial distance $R_{0}$ 
in units of $h^{-1} \ Mpc$. 
Upper panel corresponds to three
BV realizations with the same wall width $d_{w}=5h^{-1}
\ Mpc$
and three different amplitudes displayed inside the 
panel. The BV realizations of the
intermediate panel correspond to the fixed 
amplitude $(\frac {\Delta \rho}{\rho})_{max}=4$ and
the wall widths are shown inside the panel
in units of $h^{-1} \ Mpc$.
Bottom panel corresponds to 
$(\frac {\Delta \rho}{\rho})_{max}=2.5$, $d_{w}=2h^{-1}
\ Mpc$.
 
\vskip 0.5cm
\noindent
{\bf Fig. 2.} Left panels show the residual anisotropy 
$\delta_{_{R}} \times 10^{7}$
as a function of the observation angle $\psi$ (in degrees) for several 
BV realizations placed at $z=0.052$. The density parameter is 
$\Omega_{0}=0.2$.
Upper (bottom) left  panel corresponds to the same BV 
realizations as in the upper (intermediate) panel of Fig. 1. 
Right panels display the quantity
$\delta_{_{R}} \times 10^{6}$ for the same models as in
the left panels. Void centres are located at $z=2.42$.
\vskip 0.5cm
\noindent

{\bf Fig. 3.} Same as Fig. 2. 
Top panel corresponds to the realization  
$(\frac {\Delta \rho}{\rho})_{max}=4$, $d_{w}=5h^{-1}
\ Mpc$ evolving in an open universe with
$\Omega_{0}=0.2$. This realization is placed at the redshifts 
displayed inside the panel.
In the intermediate panel the realization and the redshifts 
are identical to those of the top panel, but the
background is flat.                
The bottom panel corresponds to 
the energy density 
profile of the bottom panel of Fig. 1. The background is flat. 
\vskip 0.5cm
\noindent

{\bf Fig. 4.} Density contrast $\Delta\rho/\rho$
as a function of the radial distance $R$ at redshift 2.42. 
$R$ is given in units of $h^{-1} \ Mpc$. 
The three
BV realizations have the same wall width $d_{w}=5h^{-1}
\ Mpc$
and three different amplitudes displayed inside the 
panel. The density parameter is $\Omega_{0}=0.2$.
\vskip 0.5cm
\noindent

\newpage

\begin{table}
\begin{center}
{\bf Table 1.} BV models. $\Omega_{0} =$ 0.2.\\
 \begin{tabular}{ccccccc}
\hline
$(\frac {\Delta \rho}{\rho})_{max}$ 
  &  $d_{w}$
       & $\varepsilon_{1} \times 10^{3}$ &
$\varepsilon_{2} \times 10^{2}$
& $R_{x1} \times 10^{2}$ & $R_{x2} \times 10^{2}$ & $R_{c}$ \\
       &  $(h^{-1} Mpc)$
  & & & $(h^{-1} Mpc)$  &  $(h^{-1} Mpc)$
 & $(h^{-1} Mpc)$ \\
\hline
   $2.$  &  $5.$  & $34.22$  &  $-4.29$  &  $2.60$  &  $2.32$  &  $35.9$\\
   $4.$  &  $5.$  &  $5.70$  &  $-1.44$  &  $3.63$  &  $2.09$  &  $32.7$\\
   $6.$  &  $5.$  &  $4.91$  &  $-1.36$  &  $4.31$  &  $2.06$  &  $32.4$\\
   $4.$  &  $3.$  &  $23.44$ &  $-3.21$  &  $2.70$  &  $2.32$  &  $33.3$\\
   $4.$  &  $7.$  &  $4.63$  &  $-1.33$  &  $4.21$  &  $2.05$  &  $32.8$\\
\hline
\multicolumn{7}{c}{}\\
\end{tabular}
\end{center}
\end{table}

\begin{table}
\begin{center}
{\bf Table 2.} BV models. $\Omega_{0} =$ 1.\\
 \begin{tabular}{ccccccc}
\hline
    $(\frac {\Delta \rho}{\rho})_{max}$ 
  &  $d_{w}$ & $\varepsilon_{1} \times 10^{3}$ &
$\varepsilon_{2} \times 10^{3}$
& $R_{x1} \times 10^{2}$ & $R_{x2} \times 10^{2}$ & $R_{c}$ \\
        &  $(h^{-1} Mpc)$
  & & & $(h^{-1} Mpc)$  &  $(h^{-1} Mpc)$
 & $(h^{-1} Mpc)$ \\
\hline
   $4.$  & $5.$  &  $1.43$  &  $-3.63$  &  $3.65$  &  $2.08$  &  $32.7$\\
  $2.5$  &  $2.$  &  $0.14$  & $-12.06$  &  $6.53$  &  $0.99$  &  $51.6$\\
\hline
\multicolumn{7}{c}{}\\
\end{tabular}
\end{center}
\end{table}

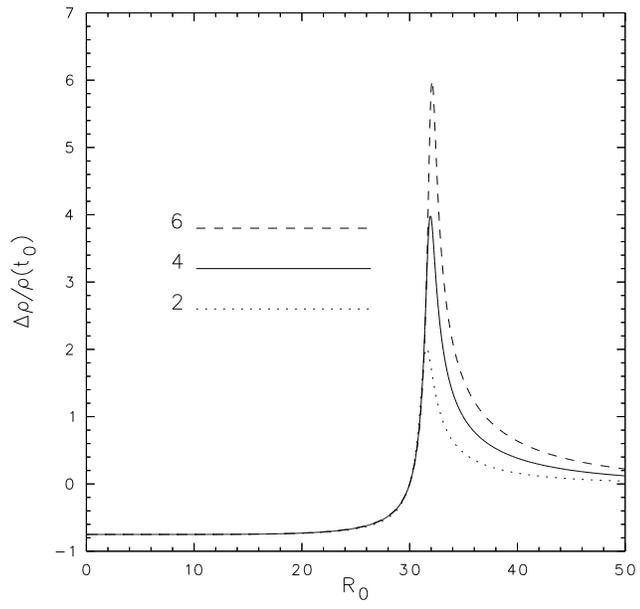
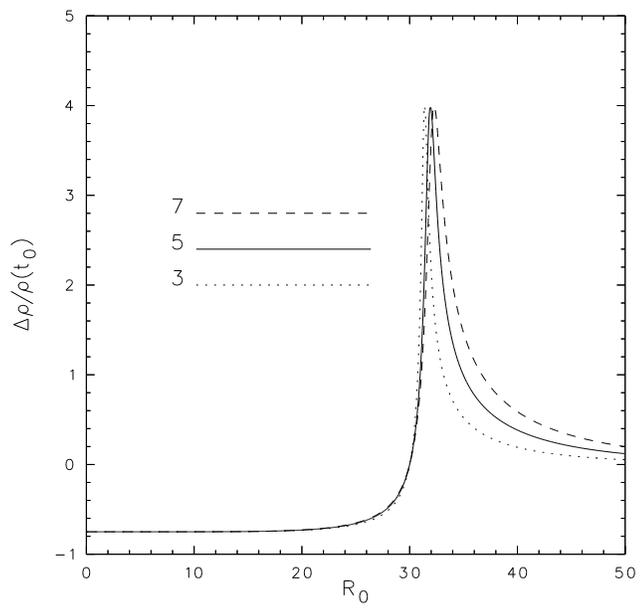
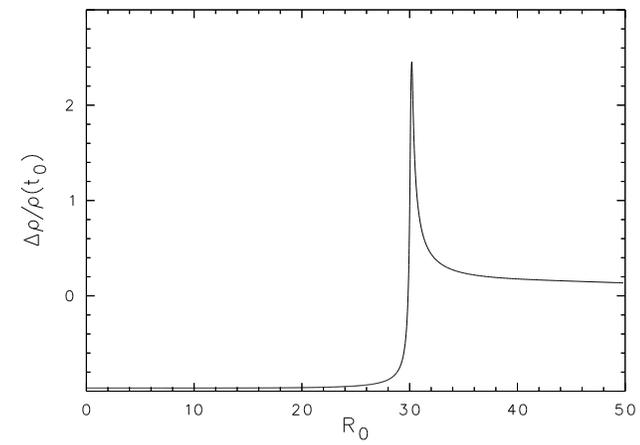
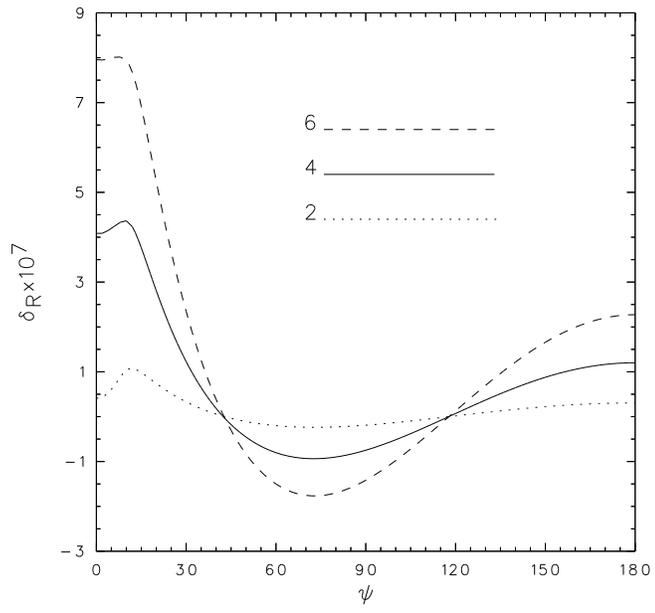
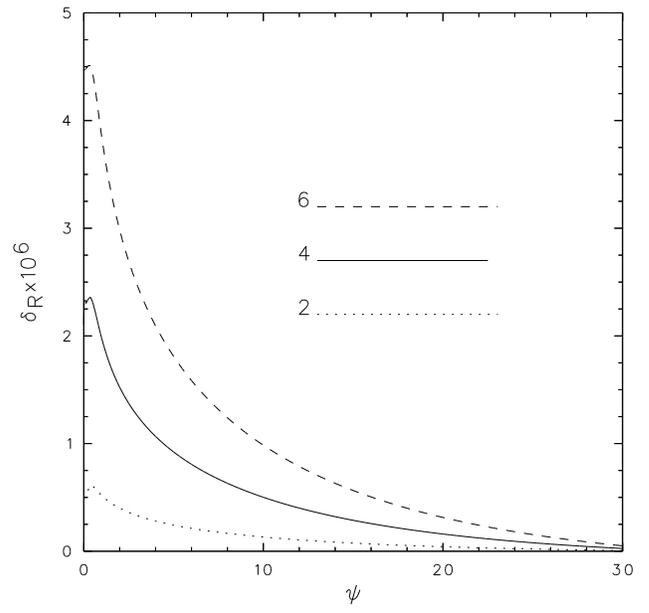
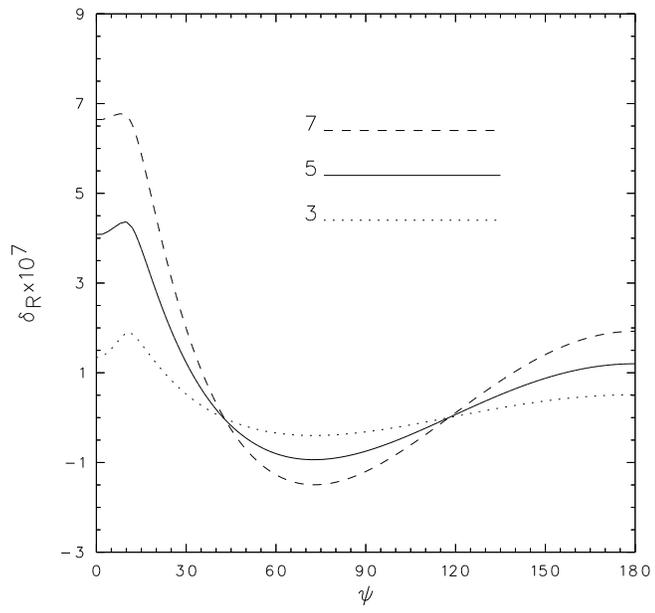
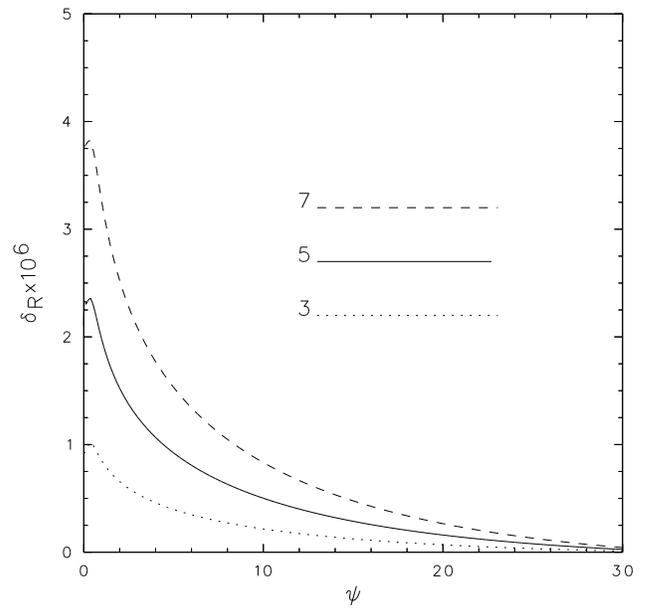
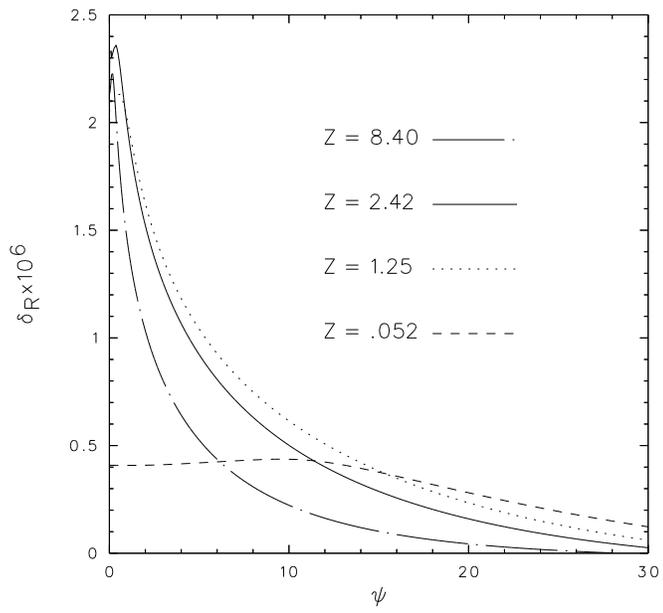
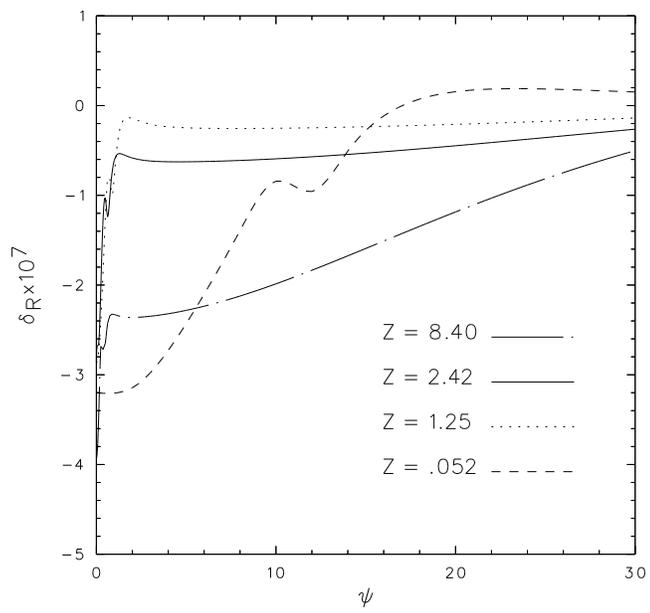
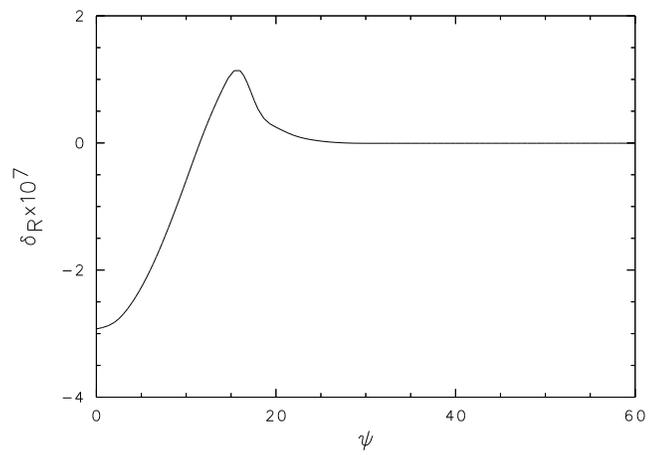
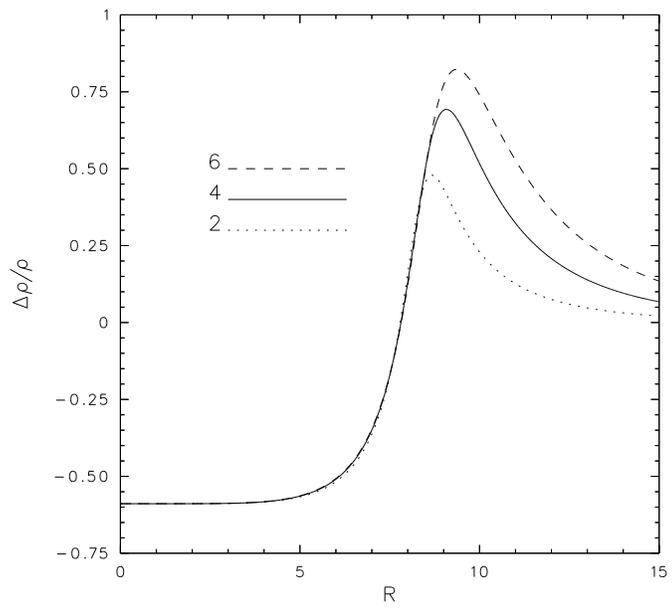
\end{document}